\documentclass[preprint, english,eqsecnum,aps,pre,showpacs]{revtex4}

\usepackage[T1]{fontenc}
\usepackage{amsmath}
\usepackage{amssymb}

\makeatletter
\usepackage{graphicx}   
\usepackage{dcolumn}    
\usepackage{bm}         
\usepackage{babel}

\begin{document}

\title{Transmission through  potential barriers with a generalised Fermi-Dirac current}

\author{J.L. ~Domenech-Garret $^{1}$}
\email{domenech.garret@upm.es}

\affiliation{$^{1}$ Dpto. F\'{\i}sica Aplicada. 
E.T.S.I. Aeron\'autica y del Espacio.\\
Univ. Polit\'{e}cnica de Madrid,  Madrid, Spain.}

\date{\today}

\begin{abstract}
We study the effect of the shape of different potential barriers on the transmitted charged current whose fermionic population  is not monoenergetic but is described by means of an energy spectrum. The generalised  Kappa  Fermi-Dirac distribution function is able to take into account both equilibrium and  non-equilibrium populations of particles by changing the kappa index. Moreover,  the corresponding current density can be evaluated using such a distribution. Subsequently, this progressive charged density arrives at the potential barrier, which is considered under several shapes, and  the corresponding transmission factor is evaluated. This permits a comparison of the effects for different barriers as well as for different energy states of the incoming current.       

\end{abstract}

\pacs{05.30.Fk; 03.65.-w; 51.10.+y}

\maketitle

\section{Introduction}
\label{intro}

Potential barriers can be found in almost any field of Physics. The usual  procedure to study  transmission considers a single particle or an ensemble with a given energy traveling through a potential barrier with a specific shape. In nuclear alpha decay \cite{Evans} is studied the probability of tunneling for a single alpha particle through the potential barrier created by the daughter nucleus and the alpha particle moving within. Moreover, incoming currents of particles with an energy distribution traveling towards a potential barrier  play an important role in calculations of the reaction rates of stellar interiors, \cite{Rolfs}. The pass through barriers of charges  are widely analysed in solid state physics, such as the study of charges with specific energy  through a periodic potential, \cite{Kittel}. Besides, concerning  thermionically emitted current of electrons, the energy of the electron population progressing to the barrier at a solid surface is no longer monoenergetic but such a current follows an energy distribution \cite{Richardson, Dushmann}. In classical works, \cite{Fowler}, assuming the equilibrium, such electron population is described by means of the Fermi-Dirac distribution. In this later reference, the value of the transmitted current is calculated through a specific barrier, the so called hill potential. The main pourpose of this work is to go further, by  considering a charged fermionic population which can  be found out of equilibrium. To achieve this goal we  build a current using the generalised Fermi-Dirac distribution function \cite{Treumann1, Treumann2}, which on taking low values of the Kappa index  is able to describe  non-equilibrium populations. The equilibrium can be recovered with the $\kappa\rightarrow\infty$ limit.  The corresponding current is studied travelling towards a barrier having different  representative shapes, namely, the step potential, the potential barrier, as well as a hill potential. Within the Conclusions section, finally, we will briefly discuss these currents but through a potential of an arbitrary shape.   

\subsection{The Kappa distribution}

The Kappa distribution based on the generalised-Boltzmann statistics is a function widely used in astrophysics \cite{Podesta,Shizgal}. An extension of such distribution into the quantum domain is the general  expression for the Fermi Dirac generalised $\kappa$-distribution function \cite{Treumann1, Treumann2}. The generalised Fermi Dirac Kappa  function, can be cast in the framework of the quantum Boltzmann equation taking into account two-particle correlation functional  containing correlational interactions of fermions. In the case of uncorrelated interactions, the two-particle correlation term  becomes the usual Fermi  two-particle correlation, \cite{Treumann1, Domenech-G5}.  At this point we should stress that the Kappa generalised Fermi Dirac distribution is not the only way to describe a fermion population departing from equilibrium but producing the same large-Kappa limit. A non equilibrium function describing such fermions could be also build using the so called relaxation time approximation, \cite{Ashcroft}. In addition, a study of \cite{Kim} concerning  the field-free equilibration of  excited nonequilibrium hot electrons in semiconductors shows the energy dependence of the relevant scattering rate plays an important role in determining the relaxation of the hot-electron distribution.  Finally, concerning again the Kappa distributions, a comprehensive study about their suitability to describe particle populations out of equilibrium can be found in \cite{Shizgal2}. This later work analyses the  dynamics towards a large Kappa limit within the physical framework of the Fokker-Plank equations.

The Kappa Fermi Dirac distribution function has been used to study the behaviour of an electron population within solid state physics from a  phenomenological point of view \cite{Domenech-G1, Domenech-G2, Domenech-G3}. Moreover, the suitability of such a generalised function describing the thermionic emission under non equilibrium conditions has been tested by means of experiments \cite{Domenech-G4}. Within the framework of the quantum Boltzmann equation above mentioned, The Kappa Fermi Dirac function might model the energy spectrum of a  fermionic gas at a temperature $T$, and can be written in the following way, \cite{Domenech-G5}, 

\begin{equation}
 f_{\kappa}( T, E) 
 =  \left[  1+ \biggl( 1 + \frac{E - \epsilon_{F}}{E_{\kappa} + \epsilon_{F}\, } \biggr)^{(\kappa+1)} \right]^{-1}
\label{eq:k-FD}
\end{equation}
\noindent
where $E$ is the fermion energy, $k_{B}$  the Boltzmann constant, and $\epsilon_{F}$ being the Fermi energy. Hereafter we label, to shorten, $E_{\kappa} = (\kappa - 3/2 )\, k_{B} T$. The above expression has a physical meaning if $\kappa >$3/2.  As expected,  for high Kappa values corresponding to equilibrium, $f_{\kappa\rightarrow\infty}( T ,E )$  recovers the FD distribution. The above generalised FD distribution  becomes  smoothed for low  Kappa values when $(E - \epsilon_{F})\ll  k_{B} T$ or $(E - \epsilon_{F})\gg  k_{B} T$. Under such conditions, \cite{Domenech-G1}, the above function can be rewritten to describe the velocity distribution of the fermion population , 

\begin{equation}
\label{fk-v}
f_{\kappa} ( |\bm{v}| ) = C_{\kappa}(T) \,  \biggl( 1\ 
+ \frac{\mu\ \bm{v}^2 - \epsilon_{F} }{E_{\kappa} + \epsilon_{F}\, } \biggr)^{-(\kappa+1)} 
\end{equation}
\noindent
Here, m being the particle mass, $\mu= m/2$, and $\bm{v}^2 =  ( v^2_{x} + v^2_{y}+ v^2_{z})$ is the velocity. The  constant $C_{\kappa}(T) $ is normalized to the  density of the electron population, \cite{Domenech-G1}. Such a distribution takes also into account the Fermi level. The suprathermal tails in the distribution develop for low $\kappa > 3/2$ values,  whereas  in the limit  $\kappa \rightarrow \infty$ the r.h.s. of  Eq. (\ref{fk-v}) reduces to a Maxwellian-like distribution. 

\section{Currents using the $\kappa$ distribution.}
\label{sec:Curr}

Throughout this work we will deal with currents flowing  along the positive $\bm{OX}$ axis.  The   current density  $J^0_{\kappa} $  parallel to the direction $\bm{u}_{x}$  is given by, 

\begin{equation}
\label{JKappa-a}
J^0_{\kappa} = \, e \,  \int^{\infty}_{v_{0x}}  \int^{\infty}_{-\infty}  \int^{\infty}_{-\infty} 
\! \!  v_{x}\ \ f_{\kappa}\ d\bm{v}    
\end{equation}
\noindent
where we set the additional superscript ''$0$'' for convenience. Here, $v_{0x}$ is a given minimum velocity along the $\bm{OX}$ axis that will be discussed later. We evaluate the current density $J_{\kappa }$ from Eq. (\ref{JKappa-a}) by  using the $\kappa$ distribution (\ref{fk-v}). 

\begin{eqnarray}
J^0_{\kappa} =\ e\  C_{\kappa} (T)\,  
\int^{+\infty}_{v_{0x}}  \!\!\!\!  v_{x} \, dv_{x}
\int^{+\infty}_{-\infty}  \!\!\!\! dv_{y} \times \nonumber \\ 
\times  \int^{+\infty}_{-\infty} \!\!\!\! dv_{z}\  \biggl( 1\ + \frac{\mu ( v^2_{x} + v^2_{y}+ v^2_{z}) - \epsilon_{F}}{E_{\kappa} + \epsilon_{F}\, } \biggr)^{-(\kappa+1)}
\end{eqnarray}

\noindent
 After integration over the z and y-directions we have;

\begin{eqnarray}
\label{JK0}
J^0_{\kappa}  =\ e\  A_{\kappa}(\epsilon_{F})\,  
   \int^{+\infty}_{v_{0x}} \!\!\!\! v_{x} \, dv_{x}\  \biggl( 1\ + \frac{\mu v^2_{x}- \epsilon_{F}}{E_{\kappa} + \epsilon_{F}\,} \biggr)^{-\kappa}
\end{eqnarray}
\noindent
where $A_{\kappa}(\epsilon_{F})$ is a term that depends on its Fermi level and the state of the fermion population through the $\kappa$ index. As we shall see, we will work in terms of the energy along the $\bm{OX}$ axis, and the final form of such integral will be later rewritten.

\subsection{Transmission coefficient and  normalisation to the incident currrent.}
\label{t-barr}
As it is well known, if we set a potential barrier in a given point with an incoming charged current, the transmission coefficient will determine the  reflection rate  of such a current. Therefore, the corrected form for the current density should include the quantum transmission coefficient $B(|\bm{p}|)$  which is a function of the momentum of the incoming particles. At this point, we take  this  transmission coefficient $B(v_x)$ along the $\bm{OX}$ axis. This coefficient is included within the integral:

\begin{equation}
\label{JTRNS}
J^B_{\kappa}= e  \int^{+\infty}_{v_{0x}} \! \! \int^{+\infty}_{-\infty}   
\! \! \int^{+\infty}_{-\infty}
\! \! v_{x}\ f_{\kappa} \ B(v_x) \ dv_{x}\ dv_{y}\ dv_{z}
\end{equation}
\noindent
Again, we label this corrected current density with the superscript ''$B$ '' for convenience.  
Using the Eq.(\ref{fk-v}) in Eq.(\ref{JTRNS}) after the integration over $v_y$ and $v_z$, its corresponding contribution  reads, 

\begin{eqnarray}
\label{JT1}
J_{\kappa} = \ e\  A_{\kappa}(\epsilon_{F})\,  
   \int^{+\infty}_{v_{0x}}  \! \!
v_{x}\  \biggl( 1\ + \frac{\mu v^2_{x} - \epsilon_{F}}{E_{\kappa} + \epsilon_{F}\, } \biggr)^{-\kappa}\  B(v_x)\ dv_{x} 
\end{eqnarray}
\noindent
The  lower integration limit, $v_{0x}$,   will then physically correspond to the velocity minimum to overcome the potential barrier. This minimum will depend on the shape of each barrier.  Usually, the normalisation constant of the  distribution function $f_{\kappa}$ entangles the information about the particle density and other physical parameters of the gas which constitute the current flowing towards the barrier. However, the exact knowledge about these quantities  cannot sometimes be achieved. Therefore, as  we are interested in the effect of the barriers on such charged currents, we will normalise the transmitted current density, Eq.(\ref{JT1}), to the incident one without the presence of the barrier, Eq.(\ref{JK0}): The norm $N$ will be $J^0_{\kappa}$. 

\begin{equation}
\label{JN}
J_{\kappa} \equiv  \frac{ J^B_{\kappa}}{N}= \frac{ J^B_{\kappa}}{J^0_{\kappa}}
\end{equation}
\noindent
Hereafter we will consider the energy of the particle distribution along the $\bm{OX}$ axis, $E= m v^2_{x}/2$. The corresponding energy minimun to overcome the potential barrier is labelled as $E_0= m v^2_{0x}/2$. Therefore, using Eqs.(\ref{JK0}) and (\ref{JT1}) in  Eq.(\ref{JN}), the normalised current density in terms of the energy finally reads,
 
\begin{equation}
\label{JNE}
J_{\kappa}= \frac{1}{N} \int^{+\infty}_{E_{0}}  
  \biggl( 1\ + \frac{E - \epsilon_{F}}{E_{\kappa} + \epsilon_{F}\,} \biggr)^{-\kappa}\  B(E)\ dE 
\end{equation}
\noindent
with
\begin{equation}
\label{Norm}
N= \int^{+\infty}_{E_{0}} 
  \biggl( 1\ + \frac{E - \epsilon_{F}}{E_{\kappa} + \epsilon_{F}\,} \biggr)^{-\kappa}\  \ dE
\end{equation}
\noindent
Here the transmission coefficient $B(E)$ will be written in terms of the energy.
Actually, if we take into account the distribution function of the incoming particles, Equation (\ref{JNE}) can be regarded to some extent as an average of  the transmission coefficient over the  $[ E_{0}, +\infty ]$  energy range.

\section{Step Potential}

The step potential  is the simplest model for a potential barrier. The physical picture would correspond to a charged current which feels  an abrupt change between two regions which have different potentials. We set the step  at $x=0$, and we label with ''$1$'' the negative region, and ''$2$'' the positive one.   Namely,
$$
V_1(x\ <\ 0)=\ 0\ \ ;\ 
V_2(x\ \geq\ 0)=\ V_0
$$
\noindent 
In this case the particles must have a minimum energy of   $E_0 = V_0$,  otherwise the  wave function within the region $2$ would decay exponentially.  Thus, we will insert this $E_0$ value within the  corresponding norm $N$, in  Eqs.(\ref{JNE}), (\ref{Norm}).

From the grounds of Quantum Mechanics, see for instance \cite{Gasiorovitz}, the corresponding barrier coefficient can be found: The particles  have a wave function which  is a linear combination of  $\Psi_1 \sim exp(\pm i p_1 x/\hbar)$ in the region $1$, with momentum  $p_1=\sqrt{2 m E}$ .  As the particles must have  energies  $E>V_0$, in the region $2$  the wave function  is written in terms of $\Psi_2 \sim exp(\pm i p_2 x/\hbar)$  with momentum   $p_2=\sqrt{2 m (E - V_0)}$.  We use the continuity of the wavefunctions and their coresponding derivatives at the $x=0$ point. After a long but straightforward calculation ,then the transmission coefficient yields,

\begin{equation}
\label{Bstep1}
B(E)=\frac{4\  p_1\  p_2}{(p_1\ +\ p_2)^2}
\end{equation}
\noindent
If we perform the change of variable $u \equiv (E-V_0)/(k_B T)$, then the barrier coefficient takes the form,

\begin{equation}
\label{Bstep2}
B(u)=\frac{ 4 \sqrt{ u k_B T}  \sqrt{ u k_B T + V_0}}{ \left(  \sqrt{u k_B T} + \sqrt{ u  k_B T+ V_0} \right) ^2 }
\end{equation}
\noindent
We will focus on the case $ k_B T/V_0 <<< 1$, then by expanding B(u) and  labelling $Q= V_{0}/k_{B} T$, we can find the simplified form  

\begin{eqnarray}
\label{Bstep-u}
B(u)=\ 4\sqrt{\frac{u}{Q}}\  \big[ 1 - 2\sqrt{u/Q}\  
- \frac{5}{2}\left( u/Q \right)+...(h.c.) \big]  
\end{eqnarray}

\noindent
Inserting this later transmission term  and the Eq.(\ref{Norm}) within the equation (\ref{JNE}),   after calculations, we finally attain,
\begin{eqnarray}
\label{Jk-STEP}
J_k =\  \frac{(2 \kappa - 1)\sqrt{\pi}}{\sqrt{2\ Q}} 
\times \biggl[ 
\frac{\Gamma(\kappa - \frac{3}{2}) \left(2 Q + 2 \kappa -3\right)^{1/2}  }{\Gamma(\kappa )}- \nonumber \\
 - \frac{2\sqrt{2}}{\sqrt{\pi Q}}   \frac{2 Q + 2 \kappa -3}{\kappa ^2- 3 \kappa+ 2} 
- \frac{15\ \Gamma(\kappa - \frac{5}{2})\left(2 Q + 2 \kappa -3\right)^{3/2} }{8\ Q\ \Gamma(\kappa )}
  \biggr] \nonumber \\ .
\end{eqnarray}

\begin{figure}[htb!]
\centerline{\hbox{ 
\includegraphics[width=8.0cm]{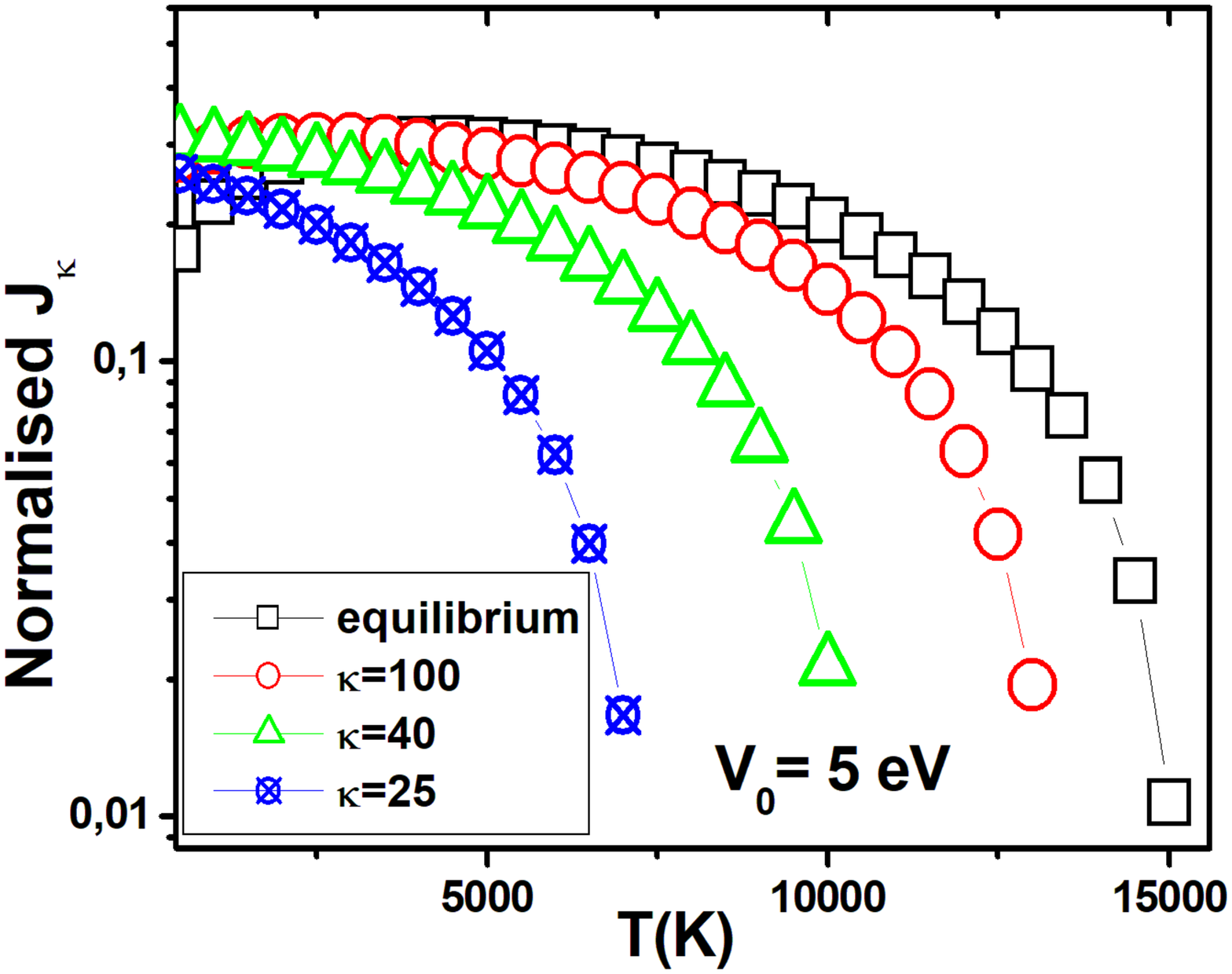} 
\includegraphics[width=8.0cm]{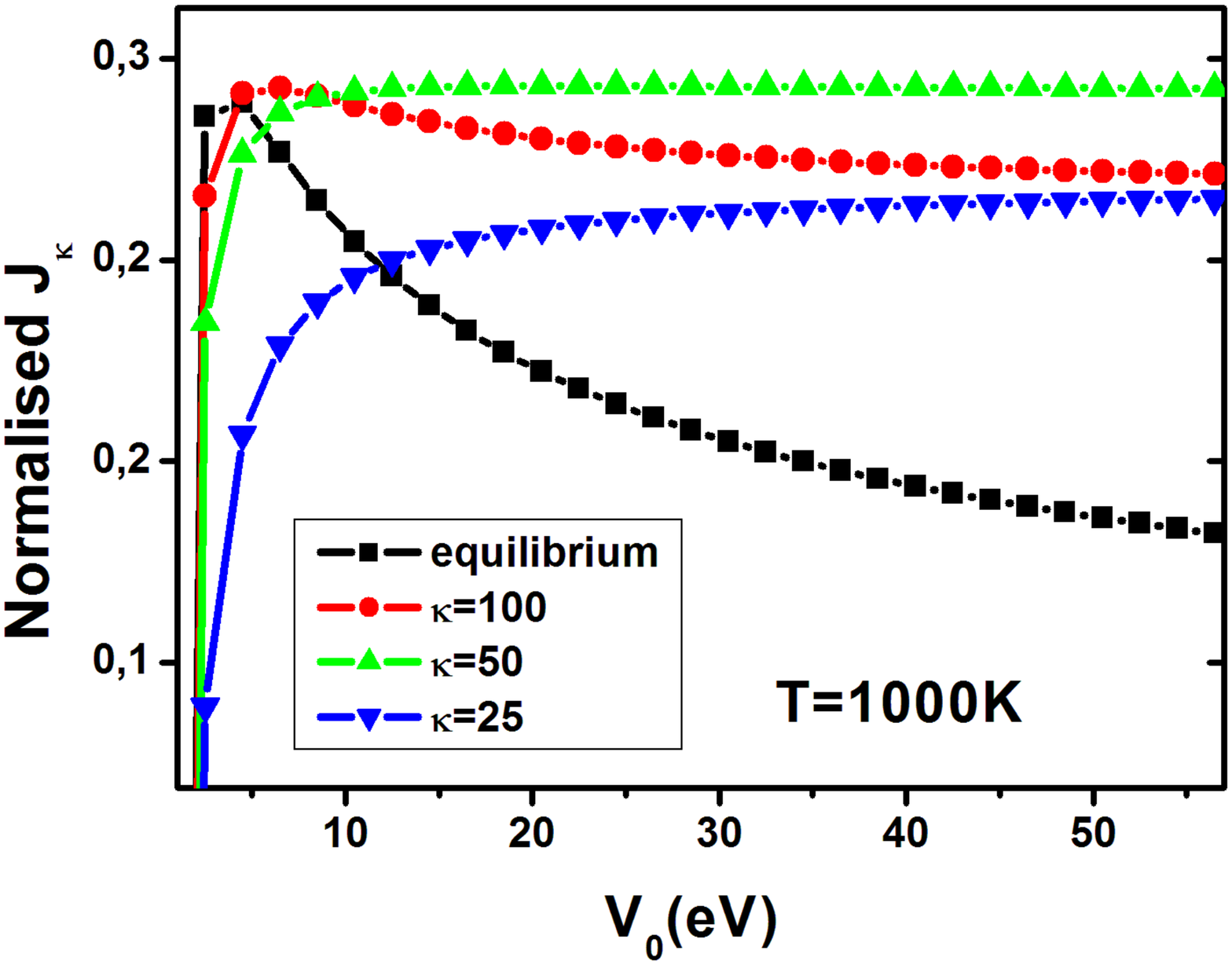}  }}
\caption{(color on line) Left-(A): The Normalised current density $J_{\kappa}$ with a  step potential as a function of temperaure for different $\kappa$ values. Right-(B): The Normalised current density $J_{\kappa}$  with a  step potential as a function of $V_0$ for different $\kappa$ values. }
\label{fig:Fig1}
\end{figure}

We can observe that the above normalised current does not depend on the mass of the incoming particles.  Using the asymptotic form of the Gamma function, \cite{Abramovitz}, in the equilibrium limit ($\kappa\rightarrow\infty$),  from the above equation we obtain,

\begin{eqnarray}
J_{\kappa\rightarrow\infty}= 4\sqrt{k_{B}T/V_0} \ \big[ \ \sqrt{\pi}/2 - \nonumber \\ 
-\ 2\sqrt{k_{B}T/V_0}- 
\frac{15\sqrt{\pi}}{8}\left( k_{B}T/V_0 \right)\  \big] 
\end{eqnarray}
\noindent
In Figure \ref{fig:Fig1}(A) is shown the evolution of the normalised current $J_{\kappa}$ as a function of the temperature. We observe the drop in the current. In all of these cases this decay is faster as the current departs from the equilibrium. Figure \ref{fig:Fig1}(B) shows the evolution of the normalised current $J_{\kappa}$ as a function of the potential $V_0$. Here we can observe the change in behaviour depending on the Kappa index: A current near equilibrium rises and then it drops slowly as $V_0$ increases. The non equilibrium currents, after the corresponding growth, saturate to an asymptotic value.      

\section{Barrier}

The physical picture of the barrier potential  would correspond to a  charged current which experiences  an abrupt change in potential  in a given width between two regions which have the same potential. We put the barrier of thickness $d$ at $x=0$, and we label ''$1$'' the negative region, ''$2$'' is the label for the positive one in which the potential changes, and ''$3$'' is the positive region which has the same potential as in region $1$. Namely, 

$$
V_1(x\ <\ 0)=\ 0\ \ ;\  
V_2(0 \leq x\ \leq\ d)=\ V_0\ \ ;\ 
V_3(x\ >\ d)=\ 0
$$ 
\noindent 
As it is well known, the particles can penetrate the barrier with the  minimum energy of   $E_0 =0$. Thus, we will insert this minimum value within Eq.(\ref{JNE}) and into the  corresponding norm $N$, Eq. (\ref{Norm}).
Following the general procedure, \cite{Gasiorovitz}, the corresponding barrier coefficient can be found: 
The particles  have a wavefunction $\Psi_1 \sim exp(\pm i p_1 x/\hbar)$ in the region $1$, and the same wavefunction in the region $3$, with  momentum  $p_1=\sqrt{2 m E}$. In  region $2$, corresponding to a barrier of heigth $V_0$, we need to distinguish between particle energies below or above $V_0$: If  $E <  V_0$ the wave function is written in terms of  $\Psi_2 \sim exp(\pm p_2 x/ \hbar)$,  with  $p_2=\sqrt{2 m ( V_0 - E )}$.  If  $E  >  V_0$ the wave function becomes complex, and it is written in terms of  $\Psi_2 \sim exp(\pm i p'_2 x/\hbar)$  with  $p'_2=\sqrt{2 m (E - V_0  )}$. After calculations, in the same way as in the previous section, the corresponding transmission coefficients are attained:

\begin{equation}
\label{Bbarr1}
B(E<V_0)=\left[ 1\ +\ \frac{ \mathrm{Sinh}^2 (p_2 d/\hbar)}{4 \dfrac{E}{V_0} \left[ 1-\dfrac{E}{V_0}\right] }\right] ^{-1}
\end{equation}
\noindent
The above expression  can be simplified if we dististinguish the energy ranges: If $ E<<<V_0 $, keeping the first order,  Eq.(\ref{Bbarr1})  becomes,
\begin{equation}
\label{Bbarr1a}
B_1(E<<<V_0)=\ 16\ \frac{E}{V_0}\  \mathrm{exp} \left[ - \frac{2\ d}{\hbar} \sqrt{2 m V_0}\ \right]
\end{equation}
\noindent
If $E \simeq V_0$,  Eq.( \ref{Bbarr1} ) can be put as,

\begin{equation}
\label{Bbarr1b}
B_2(E \simeq V_0) =\ \frac{1}{1+\dfrac{m\ V_0\ d^2}{2 \hbar^2}}
\end{equation}

The calculation of the transmission coefficient in the energy range $E >V_0$ yields,
\begin{equation}
\label{Bbarr2}
B_3(E>V_0)=\left[ 1\ +\ \frac{ \mathrm{Sin}^2 (p'_2 d/\hbar)}{4 \dfrac{E}{V_0} \left[ \dfrac{E}{V_0}-1\right] }\right] ^{-1}
\end{equation}
\noindent
This later expression for energies $E>>>V_0$ can be approximated to unity. 

\begin{figure}[htb!]
\centerline{\hbox{ 
\includegraphics[width=8.0cm]{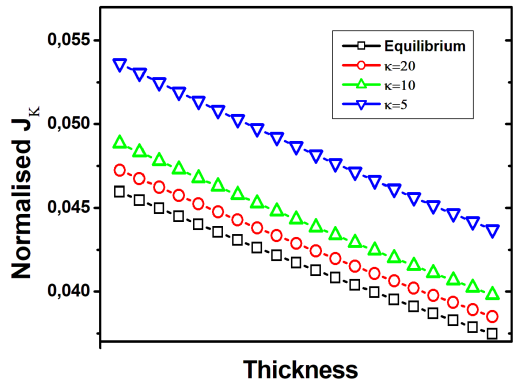} }}
\caption{(color on line)  Values of the transmited proton current density as a function of the thickness $d\in\left[10^{-10},10^{-4}\right]$m for a potential barrier of heigth $V_0= 150$ $eV$. $T=5000 K$.}
\label{fig:Fig3}
\end{figure}

As we have the expressions of $B(E)$, we will attain the corresponding normalised current density as
\begin{eqnarray}
\label{JKbarr1}
J_{\kappa} = \frac{1}{N} \int^{V_0 - \Delta}_{0}  \biggl( 1\ + \frac{E - \epsilon_{F}}{E_{\kappa} + \epsilon_{F}\,} \biggr)^{-\kappa}\   B_1(E)\  dE + \nonumber \\ + \frac{1}{N}  \int^{V_0 +\Delta}_{V_0 -\Delta}  \biggl( 1\ + \frac{E - \epsilon_{F}}{E_{\kappa} + \epsilon_{F}\,} \biggr)^{-\kappa}\   B_2(E)\  dE  +  \nonumber \\  +  \frac{1}{N} \int^{+\infty}_{V_0 +\Delta}  \biggl( 1\ + \frac{E - \epsilon_{F}}{E_{\kappa} + \epsilon_{F}\,} \biggr)^{-\kappa}\   B_3(E)\  dE \nonumber \\  
\end{eqnarray}
\noindent
Where $\Delta= \alpha V_0$ with $0<\alpha<1$. Using Eqs. (\ref{Norm}), (\ref{Bbarr1a}), (\ref{Bbarr1b}), and (\ref{Bbarr2}) within (\ref{JKbarr1}), and after integration we obtain, 

\begin{eqnarray}
\label{JKBarr2}
J_{\kappa} =  \frac{\kappa -1}{\kappa -3/2} \biggl[\  \frac{16}{V_0}\    \biggl(\  \bm{I_1}(E= V_0 - \Delta) -  \bm{I_1}(E = 0)\  \biggr)\  \mathrm{exp} \left\{ - \frac{2\ d}{\hbar} \sqrt{2 m V_0}\ \right\} + \nonumber \\ 
+ \ \left( \frac{1}{1+\dfrac{m\ V_0\ d^2}{2 \hbar^2}} - 1 \right)\  \bm{I_2}(E= V_0 + \Delta)\   
-\ \frac{1}{1+\dfrac{m\ V_0\ d^2}{2 \hbar^2}}\  \bm{I_2}(E= V_0 - \Delta) \biggr]
\end{eqnarray}
\noindent
where, to shorten the expression we introduce, 
\begin{eqnarray}
\label{I1,I2}
\bm{I_1}(E)= -\frac{E_{\kappa}^2\ - \kappa \ E_{\kappa}\  E + (\kappa -1) E^2}{ (\kappa -1) (\kappa -2)\ k_B\ T}  
\biggl( 1\ + \frac{E}{E_{\kappa}} \biggr)^{-\kappa} \nonumber \\ 
\bm{I_2}(E)=  -\frac{E_{\kappa}}{ \kappa -1} \biggl( 1\ + \frac{E}{E_{\kappa}} \biggr)^{-\kappa + 1} \nonumber 
\end{eqnarray}
\noindent
In the equilibrium limit, $\kappa\rightarrow\infty$,  Eq.(\ref{JKBarr2})  yields,
\begin{eqnarray}
\label{JBLim}
J_{\kappa\rightarrow\infty} =   16 \  \mathrm{exp} \left[ - \frac{2\ d}{\hbar} \sqrt{2 m V_0}\ \right]  \biggl[ \frac{k_B  T}{V_0}- \  \mathrm{exp} \left( - \frac{V_0 - \Delta}{k_B T} \right)  \left( \frac{k_B  T}{V_0}+\frac{\Delta}{V_0}-1 \right)       \biggr] + \nonumber \\ 
+ \ \frac{1}{1+\dfrac{m\ V_0\ d^2}{2 \hbar^2}} \biggl[   \mathrm{exp} \left( - \frac{V_0 + \Delta}{k_B T} \right) - \mathrm{exp} \left( - \frac{V_0 - \Delta}{k_B T} \right)    \biggr] -  
\mathrm{exp} \left( - \frac{V_0 - \Delta}{k_B T} \right)\  \ \  
\end{eqnarray}
\noindent
In this case, from Eq. (\ref{JKBarr2}), we see that the transmitted current is sensitive to the mass of the incoming particles. 
From the above  calculation we can perform an analysis of the behaviour of the current density as a function of the temperature. In this case we can realise a weak  dependence on the Kappa index. Hence, the current can be considered the same for all finite Kappa values. The equilibrium $\kappa \rightarrow \infty$ can only be distinguished from the others at higher temperatures. The behaviour of the $J_{\kappa}$ as a function of $V_0$ is  the same  in the case  of the  current in equilibrium  as for currents with finite Kappa values -all of them almost decay with the same exponential. Concerning the variation of  $J_{\kappa}$ as a function of the  barrier width, for an electron current the dominant term is the exponential decay with a weak dependence on the Kapppa values. The behaviour for a  proton current can be seen in Figure \ref{fig:Fig3}. Here the Kappa value becomes significant. The value of the current density which can be transmitted will be determined by the Kappa index. 

\section{Hill Potential}
\begin{figure}
\includegraphics[width=8.0cm]{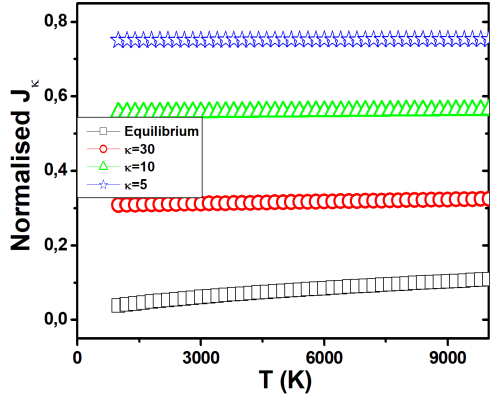} 
\caption{ (color on line) Values of the transmited proton current density as a function of the temperature through  a hill potential barrier  of heigth $V_1= 200$ $eV$, $V_0/V_1=1.3$.  \label{fig:Fig4} }
\end{figure}

\begin{figure}
\includegraphics[width=8.0cm]{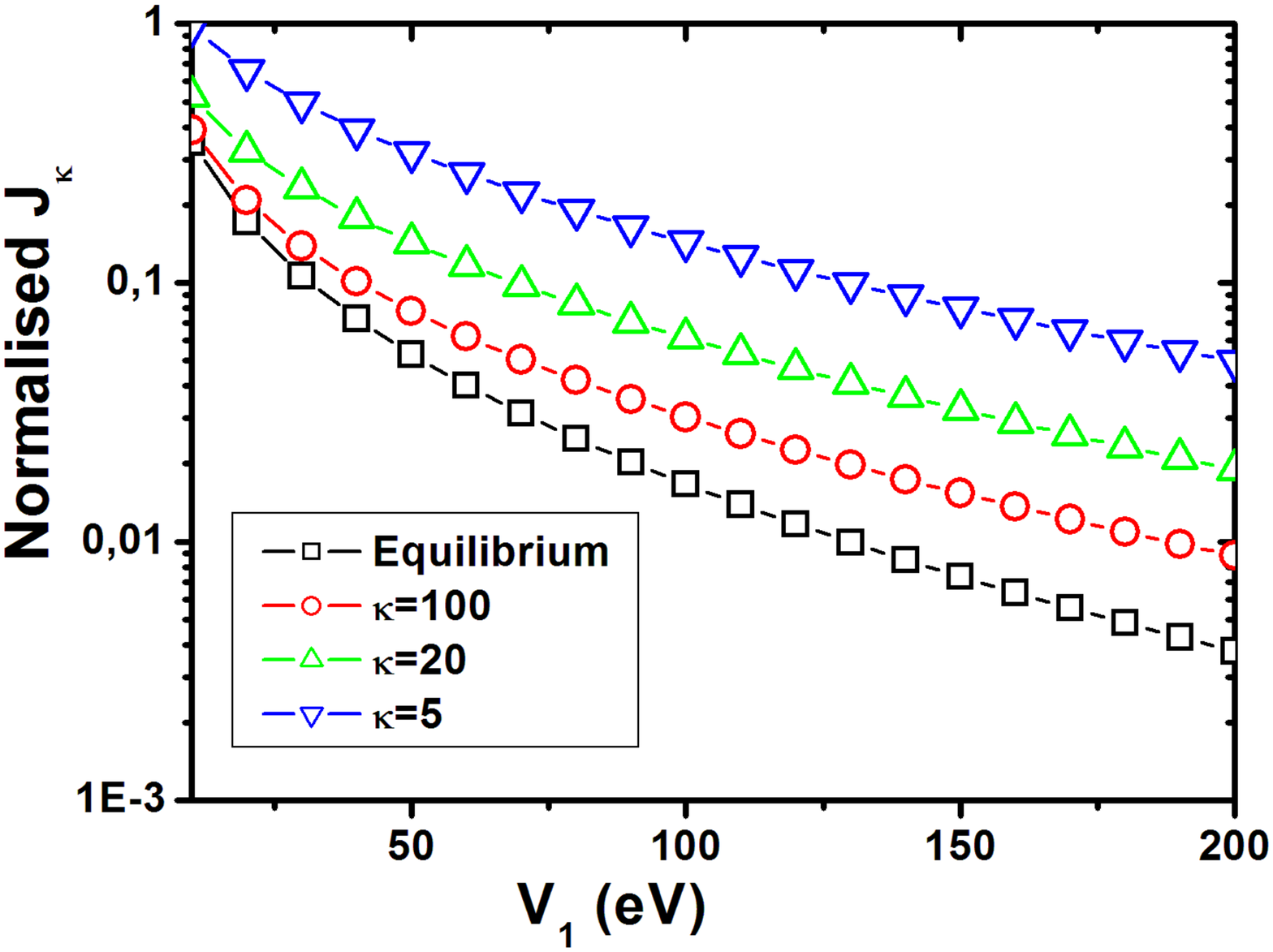} 
\caption{ (color on line) Values of the transmited proton current density as a function of $V_1$ through  a hill potential barrier  at temperature $T=5000 K$, with $V_0/V_1=3.5$.  \label{fig:Fig5} }
\end{figure}

\begin{figure}
\includegraphics[width=8.0cm]{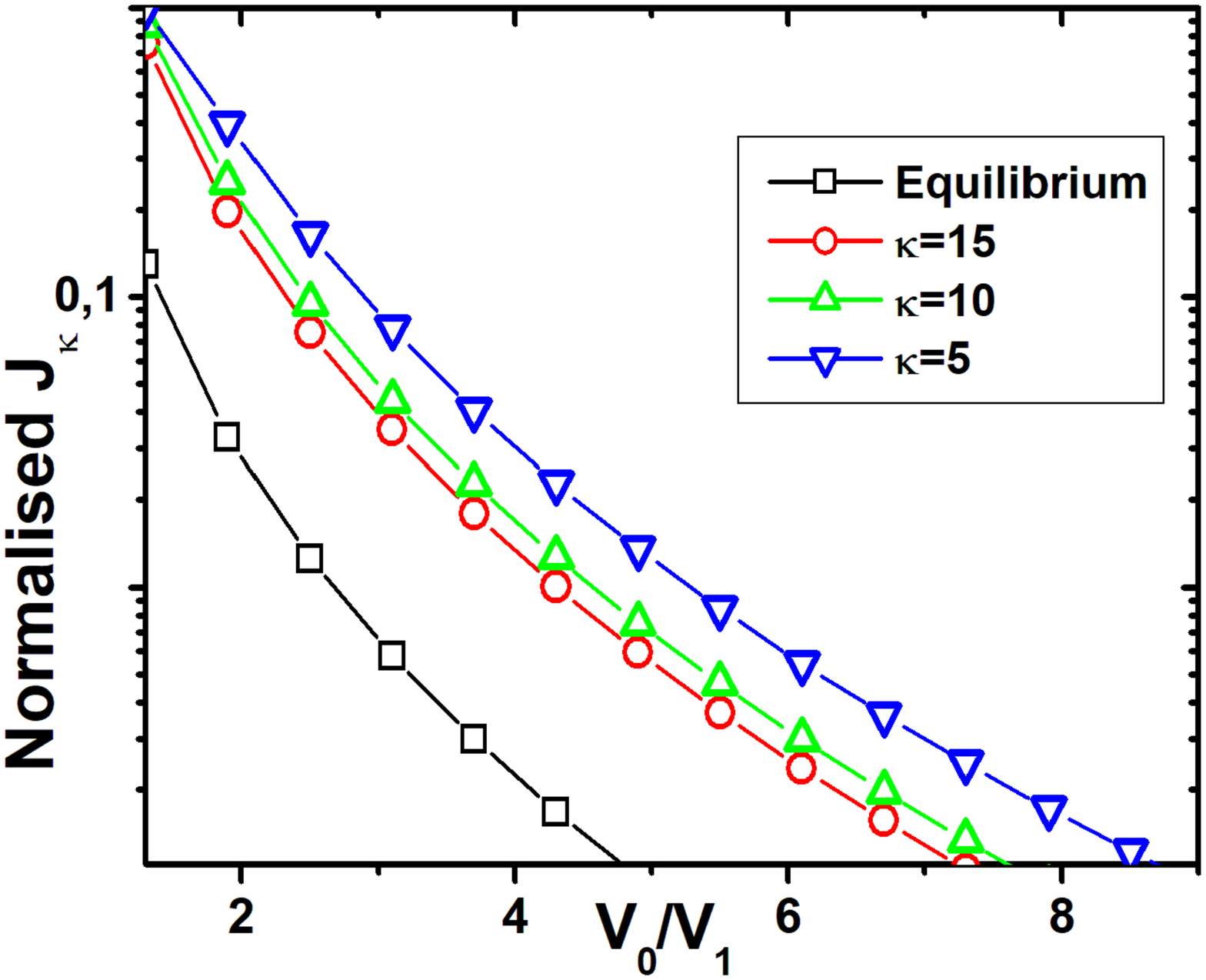} 
\caption{ (color on line) Values of the transmited proton current density as a function of $V_0/V_1$ through  a hill potential barrier at temperature $T=5000 K$, with $V_1=0.5$ $eV$.  \label{fig:Fig6} }
\end{figure}

The physical picture of the hill potential  would correspond to a  charged current which feels  
an abrupt changes at the interfaces of three regions with different potentials. The incoming current travels along a region with a ground potential. Subsequently, it penetrates  into a region of width, $d$, and potential $V_0$. Finally the current enters into a region with a  potential $V_1 < V_0$, and   greater than the ground. This kind of potential has been studied by \cite{Fowler} concerning the thermionic emission from a metal surface. Reference \cite{Fowler} takes  an incoming current obeying an equilibrium Fermi-Dirac distribution. Such a later case will correspond to the equilibrium limit in this work. We set the barrier with thickness $d$   at $x=0$, and we label ''$1$'' the negative region, ''$2$'' is the positive one in which the potential changes to $V_0$, and ''$3$'' is the positive region  after the thickness $d$, in which  the corresponding potential is  $V_1$.  

$$
V_1(x\ <\ 0)=\ 0\ \ ;\ 
V_2(0 \leq x\ \leq\ d)=\ V_0\ \ ;\  
V_3(x\ >\ d)=\ V_1\ \ (V_1 < V_0) 
$$
\noindent 
In this case the particle wavefunction cannot travel across the region $3$  unless it has a minimum energy  $E_0 =V_1$, otherwise the corresponding wavefunction within this region would become a decreasing exponential. Therefore we insert this minimum into the corresponding norm $N$, Eq.(\ref{Norm}) and  within Eq.(\ref{JNE}). The corresponding transmission coefficient  through this kind of barrier can be found  from \cite{Fowler}. For convenience we perform the following changes 

$$
u \equiv \frac{E-V_1}{k_B T}\   ;\   \  
a \equiv \frac{V_1-\epsilon_F}{k_B T}\   ;\   \
\gamma \equiv \frac{\epsilon_F}{k_B T}
$$
Hence, we can rewrite and  split Eq.(\ref{JNE})  as follows:  

\begin{eqnarray}
\label{JkV0V1-1}
J_{\kappa} = \frac{k_B T}{N} \int^{\frac{V_0-V_1}{k_B T}}_{0}  \biggl( 1\ + \frac{u\ +\ a}{\kappa-3/2+\gamma} \biggr)^{-\kappa}\   B_1(u)\  du + \nonumber \\  +  \frac{k_B T}{N} \int^{+\infty}_{\frac{V_0-V_1}{k_B T}}  \biggl( 1\ + \frac{u\ +\ a}{\kappa-3/2+\gamma} \biggr)^{-\kappa}\   B_2(u)\  du \nonumber \\  
\end{eqnarray}
\noindent
In the evaluation of the second integral $V_1+ u k_B T> V_0$, and therefore throughout this range of integration the transmission coefficient $B_2(u)$ will approximate to unity.   The transmission coefficient $B_1(u)$ in Eq.(\ref{JkV0V1-1}), within the involved range of energies,  following \cite{Fowler} and using the standard procedures, can be obtained as follows,

\begin{equation}
\label{B1-V0V1-0}
B_1(u)=\frac{16}{V_0 (V_0 - V_1)}\   \sqrt{u k_B T  \left( V_1+ u k_B T \right) }    \left( V_0- V_1- u k_B T \right) 
 \mathrm{exp} \left[ - \frac{2\ d}{\hbar} \sqrt{2 m (V_0- V_1- u k_B T )}\ \right]
\end{equation}
\noindent
The above expression with $(u k_B T/V_1)<<<$ can be expanded and simplified to,

\begin{equation}
\label{B1-V0V1-2}
B_1(u)=\frac{16}{V_0 }\   \sqrt{V_1 k_B T } \ \   
 \mathrm{exp} \left[ - \frac{2\ d}{\hbar} \sqrt{2 m (V_0- V_1)}\ \right]\  \sqrt{u}
\end{equation}
\noindent
If we label $U_M \equiv (V_0 - V_1)/k_B T$ and  $Q \equiv V_1/k_B T$, and
finally by inserting Eqs.(\ref{Norm}) and (\ref{B1-V0V1-2}) into Eq.(\ref{JkV0V1-1}) we attain,

\begin{eqnarray}
\label{JkV0V1-2}
J_{\kappa} =  \biggl(  \frac{E_{\kappa} + V_0}{E_{\kappa}+ V_1} \biggr)^{-\kappa+1}\    \times \nonumber \\  \times  \biggl[ 1  + \  \sqrt{\frac{V_1}{V_0 - V_1}}\ \  
\frac{ 16\  k_B T \left(\kappa - 1\right)}{ V_0 \left(4 \kappa ( \kappa - 2 )+ 3\right) }\ \ C_0( \kappa, U_M, Q )\ \times\  
 \mathrm{exp} \left[ - \frac{2\ d}{\hbar} \sqrt{2 m (V_0- V_1)}\ \right]\biggr] \nonumber \\  
\end{eqnarray}
\noindent
where, to shorten, the $C_0( \kappa, U_M, Q )$ term reads, 
$$ 
C_0( \kappa, U_M, Q ) \equiv 2 U_M - 2 Q + 3 -2 \kappa(2 U_M - 1) + (2 \kappa +2 Q - 3 )\  _2F_1\left[ 1\  ; \frac{1}{2}-\kappa \ ; \frac{1}{2}\  ;  \frac{-U_M}{Q+\kappa-3/2}\ \right]  
$$
and $_2F_1(a;b;c;z)$ represents the  ordinary  hypergeometric function, \cite{Abramovitz}.  
By means of the corresponding calculation in the equilibrium case, we attain,

\begin{eqnarray}
\label{JV0V1-Maxw}
J_{\kappa\rightarrow\infty} = \mathrm{exp} \left(-\frac{V_0 - V_1}{k_B T }\ \right) +  \frac{16\  k_B T }{V_0 }\   \sqrt{k_B T V_1}\ \  \mathrm{exp} \left[ - \frac{2\ d}{\hbar} \sqrt{2 m (V_0- V_1)}\ \right]\ \times
\nonumber \\  \times  \biggl[ \frac{\sqrt{\pi}}{2}\  \mathrm{Erf}\left\{\sqrt{\frac{V_0- V_1}{k_B T}}\right\}-\ \sqrt{\frac{V_0- V_1}{k_B T}}\ \mathrm{exp} \left(-\frac{V_0 - V_1}{k_B T }\ \right)\ \biggr]   
\end{eqnarray}
\noindent
where $\mathrm{Erf}(x)$ stands for the Error function, \cite{Abramovitz}. In Figure \ref{fig:Fig4} we observe the corresponding normalised current as a function of  temperature: The transmitted currents out of equilibrium (with low $\kappa$ values) are greater than the classical Fermi-Dirac case corresponding to equilibrium. In all cases, the current is almost constant with the temperature. Figure \ref{fig:Fig5} shows the normalised current as a function of the $V_1$ potential. For all Kappa values the current decays as  the potential increases, but the value of the current density  when the fermionic poulation departs from the classical Fermi-Dirac distribution becomes greater. In Figure \ref{fig:Fig6} we observe the decay of currents with different Kappa values as a function of the quotient $V_0/V_1$. In this case the decay is faster for the equilibrium ($\kappa\rightarrow\infty $) currents.  

\section{Conclusions.}
In this work, we use the generalised Fermi Dirac distribution function to build a current.  These currents have been studied  travelling through several barriers with different  representative shapes as: The step potential, the potential barrier, and also the  hill potential. The resulting currents have been later compared with those of the equilibrium limit. The transmitted current through both the step and hill potentials are quite sensitive to the shape of the potential as well as to the degree of departure from equilibrium. On the contrary, concerning the barrier, the transmission  is almost independent of the Kappa values for electron currents, and sligthy in the case of protons. A  subsequent task coming up here is to study the transmission of a Kappa current through a (smoothed) barrier of an arbitrary shape. This task could be done following the same procedure we follow along this work. Previously, the corresponding transmission coefficient could be computed as a juxtaposition of square potential  barriers of thickness $\Delta x$. The  height of potential would be suitable  to the corresponding height of the modelled potential at this $x$ point, $V(x)$. Using the WKB approximation this could be written as:  
\[ 
\mathrm{ln}\left[ B(E)\right]\approx -2\  \int_{\forall Barrier}\ dx \sqrt{2m(V(x)-E)/\hbar^2}  
\]
\noindent
and later inserting such a term into Eq.(\ref{JNE}) to obtain the transmitted current. However, it entangles to fit numerically to the modelled potential the height of $V(x)$, as well as the thickness $\Delta x$. Hence, this task turns out to be a  numerical issue, which is left for a further work.



\end{document}